\documentclass[prl,twocolumn,showpacs,preprintnumbers,amsmath,amssymb]{revtex4}
\usepackage{graphicx}
\usepackage{dcolumn}
\usepackage{bm}
\usepackage{tabularx}

  \def\S(#1,#2){{S^{#1}_{#2}}}
  \def\Su(#1,#2){{\hat{S}^{#1}_{#2}}}
  \def\Sd(#1,#2){{\check{S}^{#1}_{#2}}}
  \def\C(#1){C_{#1}}

  \def\vecr{\mbox{\boldmath $r$}}
  \def\vectau{\mbox{\boldmath $\tau$}}
  \def\vecR{\mbox{\boldmath $R$}}
    \def\vecL{\mbox{\boldmath $L$}}
    \def\vecI{\mbox{\boldmath $I$}}
       \def\vecT{\mbox{\boldmath $T$}}

  \def\tr{{\rm tr}}

  \def\gsl#1{\rlap{\slash}#1} 
   
  \def\p{\gsl p}

  \def\qq{\langle\bar qq\rangle}

  \def\epsdel(#1,#2,#3,#4){\quad(\delta_{#1#3}\delta_{#2#4}
                                 -\delta_{#1#4}\delta_{#2#3})} 
  \def\<>#1{\langle#1\rangle} 
  \def\><{\rangle\langle} 
\def\ben{\begin{equation}}
\def\een{\end{equation}}
\def\bey{\begin{eqnarray}}
\def\eey{\end{eqnarray}}
\def\ba{\begin{array}}
\def\ea{\end{array}}
\def\benmrt{\begin{enumerate}}
\def\eenmrt{\end{enumerate}}

\def\psla{p{\raise1pt\hbox{$\!\!/$}}}
\def\dsla{\partial{\raise1pt\hbox{$\!\!\!/$}}}
\def\Dsla{D{\raise1pt\hbox{$\!\!\!/$}}}
\def\xsla{x{\raise1pt\hbox{$\!\!\!/$}}}

\def\vecr{\mbox{\boldmath $r$}}

\def\jmu5{j_{\mu 5}^{(i)}(0)}
\def\jnu5{j_{\nu 5}^{(i)}(0)}

\def\qq0v{\langle0\!\mid\!{\bar q}q\!\mid\! 0\rangle}

\def\qc0f{\langle0\!\mid\!{\bar q}q\!\mid\!0\rangle_{F}}

\def\qsq0f{\langle0\!\mid\!{\bar q}\sigma_{\mu\nu}q\!\mid\!0\rangle_{F}}

\def\qgdq0f{\langle0\!\mid\!{\bar q}{\cal 
S}\gamma_{\mu}D_{\nu}q\!\mid\!0\rangle_{F}}

\def\qddq0f{\langle0\!\mid\!{\bar q}{\cal
S}D_{\mu}D_{\nu}q\!\mid\!0\rangle_{F}}

\def\3mmtm{|{\bf q}|^2}

\def\eq#1{Eq.(\ref{#1})}
\def\eqs#1#2{Eqs.(\ref{#1}) and (\ref{#2})}

\def\Ref#1{[\ref{#1}]}
\def\Refs#1#2{[\ref{#1},\ref{#2}]}

\def\p0{p_0}

\def\gam3{\mbox{\boldmath{$\gamma$}}}
\def\e0{E_{0}(s_{0},s)}
\def\e1{E_{1}(s_{0},s)}
\def\e2{E_{2}(s_{0},s)}

\def\aplt{\kern0.3333em \raise 0.2ex \hbox{$<$}%
\kern-0.8em \lower0.8ex \hbox{$\sim$}%
\kern0.3333em}
\def\aplg{\kern0.3333em \raise 0.2ex \hbox{$>$}%
\kern-0.8em \lower0.8ex \hbox{$\sim$}%
\kern0.3333em}

\begin{document}

\preprint{}
\title{Bound kaon approach for the $ppK^-$ system in the Skyrme model}

\author{Tetsuo NISHIKAWA}
\email{nishi@th.phys.titech.ac.jp} 
\affiliation{%
Department of Physics, Tokyo Institute of Technology, 2-12-1, Oh-Okayama, Meguro, Tokyo 152-8551, Japan
}
\author{Yoshihiko KONDO}
\email{kondo@kokugakuin.ac.jp} 
\affiliation{%
Kokugakuin University, Higashi, Shibuya, Tokyo 150-8440, Japan
}

\date{March 10, 2007}

\begin{abstract}
The bound kaon approach to the strangeness in the Skyrme model is applied to exploring the possibility of deeply bound $ppK^-$ states.
We derive the equation of motion for the kaon in the background of baryon number two Skyrmion expressed by the product ansatz.
Collective coordinate quantization is performed to extract the spin-singlet proton-proton state.
The numerical solution of the equation of motion shows that the kaon
can acquire large binding energy for reasonable 
proton-proton relative distances.
For this deep binding, the Wess-Zumino-Witten term plays an important role.
The kaon tends to be centered between the protons.
\end{abstract}
\maketitle

\newpage
The issue of $\bar K$ nuclear bound states has received considerable interest recently.
One of the experimental findings which triggered the boom is the report by FINUDA collaboration \cite{FINUDA}.
In the experiment at DA$\Phi$NE, the collaboration observed 
that a strong back to back correlation between $\Lambda$ and proton ($p$) from stopped $K^-$ reaction for nuclear targets is seen,
and that the invariant mass spectrum of $\Lambda$ and $p$ shows a peak.
The collaboration advocates that the observation can be interpreted as a signal of the formation of deeply bound $ppK^-$ state, whose binding energy and width are $115\,{\rm MeV}$ and $67\,{\rm MeV}$, respectively. 

This experiment is motivated by the idea proposed by Akaishi and Yamazaki (AY) \cite{AY} suggesting the existence of deeply bound $\bar K$ nuclei.
It is based on the assumption that the presence of the strong attraction in $I=0$ ${\bar K}$-nucleon ($N$) channel leads to a ${\bar K}N$ bound state,  $\Lambda(1405)$ baryon.
One may then expect that when a $K^-$ is injected in a nucleus it may attract surrounding $p$'s to form a shrunk nucleus.
The $K^-$ is bound deeply so that $K^-p\rightarrow \pi\Sigma$ absorption reaction is energetically closed, and accordingly, it has a long life in a nucleus.

However, it has not yet been established that the peak observed by FINUDA really corresponds to the state that AY proposed. 
Magas {\it et al.} \cite{magas} claimed that the peak corresponds mostly to the process $K^-pp\rightarrow\Lambda p$ followed by final-state interactions of the produced particles with the daughter nucleus.
Even if we suppose the peak to be a $ppK^-$ bound state, it is strange that it is much deeper than the original AY prediction \cite{AY}: the binding energy $48\,{\rm MeV}$ and the width $61\,{\rm MeV}$.
The recent result of Faddeev calculation \cite{gal}: the binding energy $55-70\,{\rm MeV}$ and the width $95-110\,{\rm MeV}$ are at odds with both of the result by FINUDA and AY prediction.


In this paper, we approach to the issue of deeply bound $ppK^-$ states in the bound kaon (BK) approach \cite{CK} to the Skyrme model. 
In this approach, strange baryons are described as the bound states of the $\rm SU(2)_f$ soliton (Skyrmion) and kaon.
The degeneracy of the strangeness $S=\pm1$ states is resolved owing to the Wess-Zumino-Witten (WZW) term; the $S=-1$ state is bound, while the $S=+1$ state is pushed away into the continuum.
The lowest bound state has the quantum number $L=1$, $T=1/2$, 
where $L$ is the orbital angular momentum of the kaon and $T$ the combined angular momentum and isospin, $\vecT=\vecL+\vecI$,
respectively. 
The parity of the $L=1$, $T=1/2$ state is totally positive, which is assigned to positive parity hyperons. 
A notable feature is the presence of a bound state in the negative parity state, $L=0$, $T=1/2$, which lies above the $L=1$, $T=1/2$ state. This state probably corresponds to $\Lambda(1405)$ baryon.
Whereas the constituent quark models have difficulties to describe $\Lambda(1405)$, this approach predicts the static properties of $\Lambda(1405)$ \cite{sco} as well as octet and decouplet baryons in good agreement with the empirical values.

Thus, the BK approach is a theory naturally describing both of the positive-parity hyperons and the lowest negative parity state, $\Lambda(1405)$, on the same ground.
It is also worth mentioning that this approach has no parameter once we adjust $F_\pi$ and $e$ (for their definitions, see below) to fit the $N$ and the $\Delta$ masses in $\rm SU(2)_f$ sector. 
Therefore, the ${\bar K}N$ interaction, which is a key ingredient for the study of $\bar K$ nuclei, is unambiguously determined.
In these respects, it is of great significance to investigate the issue of the exotic nuclei such as $ppK^-$, $ppnK^-$, $pnnK^-$ and so on in the context of the BK 
approach.

For the existence of nuclear $\bar K$ bound states, it is necessary that $\bar K$
gains sufficiently large binding energy in nuclei.
If such nuclei exist, the nuclear components rearrange themselves under the influence of the strong attraction in $I=0$ $\bar K$$N$.
Then the nuclear part in $\bar K$ nuclei is excited relative to the original nuclear system. Therefore, the energy gained by $\bar K$ must be large enough to compensate the energy loss of nuclear component and to deeply bind the total system.  
In this study, in order to investigate the behavior of the kaon which couples with two protons, we derive the equation of motion for the kaon field fluctuating around baryon number $B=2$ Skyrmion. 
The energy of the kaon and its dependence on the relative distance between two Skyrmions obtained from the equation of motion will tell us whether the deeply bound $ppK^-$ state can exist or not. If we know such a state can be realized, we aim at clarifying its structure and the mechanism responsible for the deep binding.

Let us begin with showing how the $K^-$ coupled to $pp$ is described in the BK approach to the Skyrme model.
We consider two Skyrmions fixed at positions with the relative distance, $R$,
and assume the presence of the kaon field fluctuating around the Skyrmions.
The equation of motion for the kaon in the background field of the Skyrmions is then derived, from which we know the behavior of the $K^-$ coupled to two protons.

The action of the Skyrme model is given by
\bey
\Gamma&=&\int d^4 x\left\{\frac{F_\pi^2}{16}\tr(\partial_\mu U^\dagger\partial^\mu U)
\right.\cr
&&\left.
+\frac{1}{32e^2}\tr\left[\partial_\mu UU^\dagger,\partial_\nu UU^\dagger\right]^2
+\tr M(U+U^\dagger-2)\right\}
\cr&&+\Gamma_{\rm WZW},
\label{action}
\eey
where $U$ is the chiral SU(3) field built out of the eight NG boson and $M$ the mass matrix. $\Gamma_{\rm WZW}$ is the WZW anomaly action \cite{witten}.
We assume the following ansatz for $U$, 
\ben
U=U(1)U_KU(2),
\label{ansatz}
\een
where $U(1)$ and $U(2)$ are the fields of the Skyrmions located at
$\vecr_1=\vecr-{\vecR}/{2}$ and $\vecr_2=\vecr+{\vecR}/{2}$, respectively.
Their explicit expressions are as follows,
\ben
U(i)=\left(\begin{array}{cc}
u(i)&\bf{0}\\
\bf{0}&0
\end{array}\right),\,\,
u(i)=e^{iF(r_i)\vectau\cdot{\hat\vecr_i}},\,\, (i=1,2),
\een
where $r_i=|\vecr_i|$, ${\hat\vecr_i}=\vecr_i/r_i$, $F(r)$ is the Skyrmion profile function and $\vectau$ the Pauli matrices.
$U_K$ is the kaon field given by
\bey
U_K=e^{i\hat{K}},\,\,
\hat{K}=\sqrt{2}\frac{2}{F_\pi}\left(
\begin{array}{cc}
\bf{0}  & K  \\
K^\dagger  & \bf{0}  
\end{array}
\right),\,\,
K=\left(
\begin{array}{c}K^+ \\ K^0\end{array}\right)
\eey
Each Skyrmion is rotated in the space of SU(2) collective coordinate $A_1$ or $A_2$
as
\ben
u(1)\rightarrow A_1u(1)A^\dagger_1,\quad u(2)\rightarrow A_2u(2)A^\dagger_2.
\label{rotate}
\een

By substituting the ansatz, \eq{ansatz}, with the replacement \eq{rotate} into the action, \eq{action}, we obtain the Lagrangian for the kaon field in the presence of the background $B=2$ Skyrmion.
After expanding the Lagrangian up to the seceond order in $K$
and neglecting $O(N_c^{-1})$ terms ($N_c$ denotes the number of colors.), we obtain
\begin{eqnarray}
{\cal L}
&=&(\partial_0K)^\dagger \partial_0K+K^\dagger D_jD_jK
\cr&&-{1\over8}K^\dagger K\Bigg\{-\tr(\partial_j U_{BB}^\dagger\partial_j U_{BB})
\cr&&+{1\over e^2F_\pi^2}\tr[\partial_j U_{BB}U_{BB}^\dagger,\partial_i U_{BB} U_{BB}^\dagger]^2\Bigg\}
\cr&&+{1\over e^2F_\pi^2}\Bigg\{
2K^\dagger D_j[D_i K\tr(A_j A_i)]
\cr&&+{1\over2}(\partial_0K)^\dagger \partial_0K\tr(\partial_j U_{BB}^\dagger\partial_j U_{BB})
\cr&&+{1\over2}K^\dagger D_j[D_jK\tr(\partial_iU_{BB}^\dagger\partial_iU_{BB})]
\cr&&-6K^\dagger D_j([A_i,A_j]D_iK)\Bigg\}
-m_K^2K^\dagger K
\cr&&-{iN_c\over F_\pi^2}[K^\dagger B^0 \partial_0K
-(\partial_0K)^\dagger B^0K]
,
\label{lag}
\end{eqnarray}
where $U_{BB}$ represents the product of rotating solitons,
$U_{BB}=A_1u(1)A^\dagger_1A_2u(2)A^\dagger_2$.
The covariant derivative $D_\mu$ is defined by $D_\mu K=\partial_\mu K+V_\mu K$ and $V_\mu=[L_\mu(1)+R_\mu(2)]/2$, $A_\mu=[L_\mu(1)-R_\mu(2)]/2$,
where
$L_\mu(1)=A_1u^\dagger(1)A_1^\dagger\partial_\mu[A_1u(1)A_1^\dagger]$ 
and $R_\mu(2)=A_2u(2)A^\dagger_2\partial_\mu[A_2u^\dagger(2)A^\dagger_2]$.
In \eq{lag}, the last term comes from the WZW term and 
$B^0$ is the time component of the baryon number current, $B^\mu$.
Here we note that the Lagrangian, \eq{lag}, has the same form as that for $B=1$ Skyrmion \cite{CK} but the background Skyrmion is expressed by the product of $B=1$ Skyrmions, instead of the single $B=1$ Skyrmion.

Our next task is to perform the collective coordinate quantization, 
and project the rotation of each Skyrmion onto the relevant spin-isospin state.
This procedure is as follows.
First, we rewrite the Lagrangian, \eq{lag}, in terms of the adjoint matrix defined by 
$D_{ij}(A)=\tr(\tau_iA\tau_jA^\dagger)/2$
with $A$ being a collective coordinate.
The matrix $D_{ij}(A)$ is known to be represented by the rotation matrix of a rank-1 tensor.
Namely, choosing the spherical basis, $\{-(1-2i)/\sqrt{2},3,(1+2i)/\sqrt{2}\}$, one has,   
$D_{ij}(A)={\cal D}^{1}_{MM'}(\Omega)$,
where ${\cal D}$ is the rotation matrix and $\Omega$ denotes the Euler angles.
On the other hand, the wave function of the nucleon with the third component of isospin $I_3$ and that of spin $J_3$ is also expressed by rotation matrix:  
$|N_{I_3,J_3}\rangle=(-1)^{I_3+1/2}D^{1/2}_{-I_3 J_3}(\Omega)/(2\pi)$.
Then, the projection of the Skyrmions onto physical two nucleon states
is performed by sandwiching the Lagrangian, \eq{lag}, with two nucleon states and integrating the Euler angles.
Here we project the rotation onto the spin-singlet proton-proton state.
Thus we consider
\bey
{\cal L}_{ppK}\equiv\int d\Omega\langle \Psi|{\cal L}|\Psi\rangle,
\label{lagKpp}
\eey
where $|\Psi\rangle$ is the wave function corresponding to the spin-singlet proton-proton state given by
\ben
|\Psi\rangle=
\frac{1}{\sqrt{{\cal N}}}(|N(1)_{\frac{1}{2},\frac{1}{2}}N(2)_{\frac{1}{2},\frac{-1}{2}}\rangle
-|N(1)_{\frac{1}{2},\frac{-1}{2}}N(2)_{\frac{1}{2},\frac{1}{2}}\rangle),
\een
with $\cal N$ being the normalization constant.
\eq{lagKpp} is the Lagrangian for the kaon coupled to two protons.

Now, we derive the equation of motion for the kaon from the Lagrangian, \eq{lagKpp}.
First, we average the direction of the line joining the two Skyrmions:
we put $\vecR=(R\sin\alpha\cos\beta, R\sin\alpha\sin\beta,R\cos\alpha)$ and integrate the angles $\alpha$ and $\beta$,
\ben
\bar{\cal L}_{ppK}=\frac{1}{4\pi}\int_0^{\pi}d\alpha\int_0^{2\pi}d\beta\sin\alpha {\cal L}_{ppK}.
\label{avlag}
\een
Then the background field becomes spherical, which allows us to set the kaon field as
\bey
K(\vecr,t)=k(r,t)Y_{lm}(\theta,\phi)
\label{kansatz}
\eey
with $Y_{lm}(\theta,\phi)$ the spherical harmonics.
This ansatz, \eq{kansatz}, is substituted into the Lagrangian, \eq{avlag},
and the $\theta$- and $\phi$-integrations are done.
Up to this step, quite long and involved calculations are needed. 
After the above procedure, the variation with respect to $k(r,t)$ yields
\bey
&&
\Big[-\bar{f}(r;R)\frac{d^2}{dt^2}-2i\bar{\lambda}(r;R)\frac{d}{dt}
\cr&&
-m_K^2-\bar{V}_{\rm eff}(r;R,L)+\hat{\cal O}\Big]k(r,t)=0,
\label{eom}
\eey
where
\bey
\hat{\cal O}&=&
c_1(r;R)\frac{\partial}{\partial r}+c_2(r;R)\frac{\partial^2}{\partial r^2}
.
\label{ohat}
\eey
In \eqs{eom}{ohat}, $m_K=495\,{\rm MeV}$ is the kaon mass
and the coefficients, $\bar{f}(r;R)$, $\bar{\lambda}(r;R)$, $\bar{V}_{\rm eff}(r;R,L)$ and $c_i(r;R)$ are written in terms of $F(r_i)$.
Their explicit forms are quite lengthy and are not particularly instructive.
Therefore we do not display them here.
Let us expand the field $k(r,t)$ in terms of its eigenmodes:
\bey
k(r,t)&=&
\sum_n\left[k_n(r)e^{i\omega_nt}a_n^\dagger
+\tilde{k}_n(r)e^{-i\tilde{\omega}_nt}b_n\right],
\label{modeexpand}
\eey
where $a_n$ and $b_n$ are the annihilation operators for the strangeness $S=\mp 1$ states, respectively.
Substituting \eq{modeexpand} into \eq{eom}, we find the eigenmodes satisfy
\bey
\Big[\bar{f}(r;R)\omega_n^2-m_K^2-V_{K}^{(-)}(r;\omega_n,L,R)+\hat{\cal O}\Big]k_n(r)=0,\,\,
\label{eoms-1}\\
\Big[\bar{f}(r;R)\tilde{\omega}_n^2-m_K^2-V_{K}^{(+)}(r;\tilde{\omega}_n,L,R)+\hat{\cal O}\Big]\tilde{k}_n(r)=0\,\,
\label{eoms+1},
\eey
where
\bey
&&V_{K}^{(\mp)}(r;\omega,L,R)=V_{WZW}^{(\mp)}(r;\omega,R)+\bar{V}_{\rm eff}(r;L,R),\label{pot}\cr
&&V_{WZW}^{(\mp)}(r;\omega,R)=\mp2\bar{\lambda}(r;R)\omega.
\label{potWZ}
\eey
\eqs{eoms-1}{eoms+1} are the equation of motions for $S=-1$ and $S=+1$ states, respectively.

\begin{figure}[t]
\begin{center}
\rotatebox{-90}
{\includegraphics[width=6.3cm,keepaspectratio]{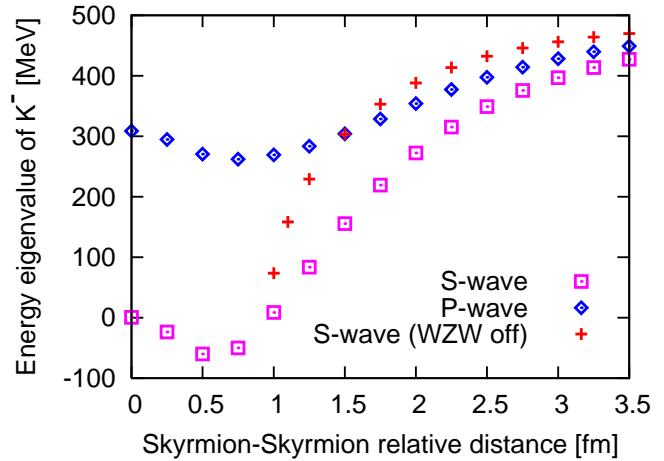}}
\caption{The energy eigenvalue, $\omega$, of s/p-wave $K^-$ as a function of the Skyrmion-Skyrmion relative distance, $R$. The crosses represent the results for the s-wave $K^-$ when the Wess-Zumino-Witten term is switched off (for $R<1\,{\rm fm}$ the kaon is unbound). 
}
\label{omega}
\end{center}
\end{figure}
\begin{table}
\begin{center}
\begin{tabular}{p{3cm}p{3cm}p{3cm}}
\hline
\multicolumn{1}{p{2.3cm}}{\hspace{.7cm}$R$ (fm)} &\multicolumn{1}{p{2.5cm}}{\hspace{.5cm}$\omega_{L=0}$\,(MeV)} & \multicolumn{1}{p{2.5cm}}{\hspace{.5cm}$\omega_{L=1}$\,(MeV)} \\
 \hline\hline
\multicolumn{1}{c}{1.0}    & \multicolumn{1}{c}{\quad9}          & \multicolumn{1}{c}{269}\\
 \multicolumn{1}{c}{1.5}    & \multicolumn{1}{c}{156}          & \multicolumn{1}{c}{304}\\
 \multicolumn{1}{c}{2.0}     & \multicolumn{1}{c}{272}          & \multicolumn{1}{c}{354}\\
 \multicolumn{1}{c}{2.5}     & \multicolumn{1}{c}{349}          & \multicolumn{1}{c}{397}\\
  \multicolumn{1}{c}{}   &\multicolumn{1}{c}{(367)}      &\multicolumn{1}{c}{(153)}\\
 \hline
\end{tabular}
\end{center}
\caption{Energy eigenvalue of s-wave $K^-$ ($\omega_{L=0}$) and that of p-wave ($\omega_{L=1}$) for four cases of the relative distance of the two Skyrmions: $R=$1.0, 1.5, 2.0, 2.5 fm.
For comparison, the values for $B=1$ \cite{sco} are also shown in the last low.}
\label{tableomega}
\end{table}
\begin{figure}
\begin{center}
\rotatebox{-90}
{\includegraphics[width=6.3cm,keepaspectratio]{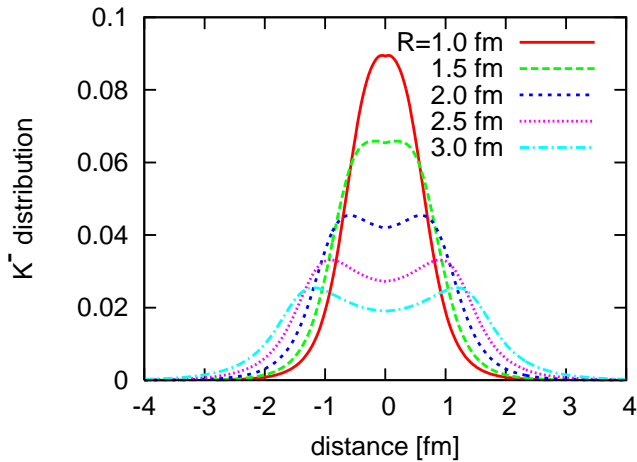}}
\caption{s-wave $K^-$ distribution (normalized wave function $k(r)$ scaled with $eF_\pi$) for the relative distance of the two Syrmions, $R=1.0$, $1.5$, $2.0$, $2.5$ and $3.0\,{\rm fm}$. The horizontal axis is the distance from the origin.}
\label{wfs}
\end{center}
\end{figure}
We solved numerically the equation of motion, \eq{eoms-1}.
Figure.\ref{omega} exhibits the obtained energy eigenvalue of $K^-$, $\omega$, as a function of the Skyrmion-Skyrmion relative distance, $R$.
We find that the lowest-lying mode is the s-wave and that the p-wave lies above the s-wave.
Note that this order is natural but different from the case of $B=1$, where
the lowest-lying mode is p-wave, as mentioned above.
In Table \ref{tableomega}, the $K^-$ energy eigenvalues for several cases of $R$ are displayed. For comparison, the values for $B=1$ \cite{sco} are also shown.
Remarkably, the s-wave $K^-$ is bound much deeper than that for $B=1$ and it can acquire huge binding energy, several hundred MeV.

Considering the $R$-dependence of the kaon's energy in Figure \ref{omega}, let us discuss the possible distance between two protons in the $ppK^-$ system.
For smaller distance, {\it i.e.}  $R\aplt 1.0\,{\rm fm}$, the binding of the kaon is extremely strong. In this region, however, the repulsive nucleon-nucleon interaction dominates over the attractive interaction between kaon and two protons.
On the other hand, for $R\aplg 2.5\,{\rm fm}$, the binding of the s-wave kaon becomes looser than that bound to $B=1$ Skyrmion, whose energy is $367\,{\rm MeV}$.
Hence, roughly speaking, if deeply bound $ppK^-$ states exist, the average distance between two protons may be in the range $1.0\,{\rm fm}\aplt R\aplt 2.5\,{\rm fm}$.

Next, we consider the possible structure of the $ppK^-$ system.
In Figure \ref{wfs},
we plot the distribution of $K^-$ in s-wave (normalized kaon wave function $k(r)$) 
for four values of $R$ in the range discussed above: $R=1.0$, $1.5$, $2.0$ and $2.5\,{\rm fm}$.
At $R=1.0\,{\rm fm}$, it can be seen that
the kaon is localized in the narrow region between the two Skyrmions.
The binding of the kaon in this case is extremely strong (see Table \ref{tableomega}).
At larger distance, $R\aplg 2.5\,{\rm fm}$, the $K^-$ distribution has a shape similar to those characteristic to molecular orbital states.
Namely, the kaon has large probability to stay near the positions of the two Skyrmions.
The binding of the kaon is much weaker than the above case.
At the intermediate distances, $R=1.5\sim2.0\,{\rm fm}$, the kaon stays in the region between the two Skyrmions on average.
Even for these values of $R$, which are close to the average inter nucleon-nucleon distance in normal nuclei,
the binding of the kaon is still deep: the binding energy exceeds $200\,{\rm MeV}$ (see Table \ref{tableomega}).  

Finally, let us look at the role of the WZW term.
It is known that in $B=1$ sector the existence of the WZW term leads to various important results.
The Skyrme Lagrangian without the term possesses a fictitious symmetry
that forbids some processes allowed in QCD.
The WZW term is required to break the symmetry \cite{witten}. 
In addition, for odd $N_c$, the WZW term ensures that the quantized Skyrmion has a half-odd spin and thus behaves as a fermion.  
The effect of the WZW term goes beyond these rules. 
In the BK approach, the WZW term gives an effective attractive contribution which is crucial for obtaining the correct values of the masses of ground state hyperons.
In particular, without the term, the s-wave bound state of a kaon and a Skyrmion, which corresponds to $\Lambda(1405)$, does not exist.
Also in the present $B=2$ case, the role of the WZW term is revealed to be important.
In the equation of motion there exist two terms, $V_{WZW}^{(-)}$ and $\bar{V}_{\rm eff}$ in \eq{pot}, which effectively play a role of the potential acting on the kaon. Among them, it is $V_{WZW}^{(-)}$ that originates from the WZW term.
In order to see the effects of the WZW term, we switched off $V_{WZW}^{(-)}$.
The result for the s-wave $K^-$ is shown in Figure \ref{omega} with crosses.
One observes that the WZW term additionally gives a substantial attractive contribution to the binding of the kaon even for relatively large $R$,
which might be crucial for the existence of the deeply bound $ppK^-$ state.

In summary, we have applied the Skyrme model to a study of the $ppK^-$ system.
We derive kaon's equation of motion in the background of $B=2$ Skyrmion expressed by the product ansatz.
Collective coordinate quantization is performed to extract the relevant spin-isospin state of the nucleons.
Numerical solution of the equation shows that the lowest-lying mode is s-wave.  The s-wave $K^-$ acquires binding energy more than 200MeV even for the relatively large proton-proton distances and is localized in the region between the protons.
The obtained results may support the suggestion for the existence of the deeply bound strange dibaryon.
To draw a more definite answer to this issue, we must solve the radial motion of the Skyrmions and estimate the binding energy of the total system, which will be reported elsewhere. 

The authors would like to thank Prof. O.~Morimatsu, Prof. M.~Oka, Prof. A.~Dote, Prof. Y.~Kanada-En'yo and Prof. Y.~Akaishi for useful suggestions and discussions.
This work was supported in part by the 21st Century COE Program at Tokyo
Institute of Technology ``Nanometer-Scale Quantum Physics'' from the
Ministry of Education, Culture, Sports, Science and Technology, Japan.

\def\Ref#1{[\ref{#1}]}
\def\Refs#1#2{[\ref{#1},\ref{#2}]}
\def\npb#1#2#3{{Nucl. Phys.\,}{\bf B{#1}},\,#2\,(#3)}
\def\npa#1#2#3{{Nucl. Phys.\,}{\bf A{#1}},\,#2\,(#3)}
\def\np#1#2#3{{Nucl. Phys.\,}{\bf{#1}},\,#2\,(#3)}
\def\plb#1#2#3{{Phys. Lett.\,}{\bf B{#1}},\,#2\,(#3)}
\def\prl#1#2#3{{Phys. Rev. Lett.\,}{\bf{#1}},\,#2\,(#3)}
\def\prd#1#2#3{{Phys. Rev.\,}{\bf D{#1}},\,#2\,(#3)}
\def\prc#1#2#3{{Phys. Rev.\,}{\bf C{#1}},\,#2\,(#3)}
\def\prb#1#2#3{{Phys. Rev.\,}{\bf B{#1}},\,#2\,(#3)}
\def\pr#1#2#3{{Phys. Rev.\,}{\bf{#1}},\,#2\,(#3)}
\def\ap#1#2#3{{Ann. Phys.\,}{\bf{#1}},\,#2\,(#3)}
\def\prep#1#2#3{{Phys. Reports\,}{\bf{#1}},\,#2\,(#3)}
\def\rmp#1#2#3{{Rev. Mod. Phys.\,}{\bf{#1}},\,#2\,(#3)}
\def\cmp#1#2#3{{Comm. Math. Phys.\,}{\bf{#1}},\,#2\,(#3)}
\def\ptp#1#2#3{{Prog. Theor. Phys.\,}{\bf{#1}},\,#2\,(#3)}
\def\ib#1#2#3{{\it ibid.\,}{\bf{#1}},\,#2\,(#3)}
\def\zsc#1#2#3{{Z. Phys. \,}{\bf C{#1}},\,#2\,(#3)}
\def\zsa#1#2#3{{Z. Phys. \,}{\bf A{#1}},\,#2\,(#3)}
\def\intj#1#2#3{{Int. J. Mod. Phys.\,}{\bf A{#1}},\,#2\,(#3)}
\def\sjnp#1#2#3{{Sov. J. Nucl. Phys.\,}{\bf #1},\,#2\,(#3)}
\def\pan#1#2#3{{Phys. Atom. Nucl.\,}{\bf #1},\,#2\,(#3)}
\def\app#1#2#3{{Acta. Phys. Pol.\,}{\bf #1},\,#2\,(#3)}
\def\jmp#1#2#3{{J. Math. Phys.\,}{\bf {#1}},\,#2\,(#3)}
\def\cp#1#2#3{{Coll. Phen.\,}{\bf {#1}},\,#2\,(#3)}
\def\epjc#1#2#3{{Eur. Phys. J.\,}{\bf C{#1}},\,#2\,(#3)}
\def\mpla#1#2#3{{Mod. Phys. Lett.\,}{\bf A{#1}},\,#2\,(#3)}
\def\etal{{\it et al.}}

\end{document}